\documentclass[aoas,MSNbibl,nameyear,seceqn,dvips]{arximspdf}
\usepackage{graphicx}
\doi{10.1214/12-AOAS541} 
\volume{6}
\issue{3}
\pubyear{2012}
\firstpage{950}
\lastpage{976}

\makeatletter

\newcommand{\underset}[2]{\mathop{#2}_{#1}}

\newtheorem{theorem}{Theorem}[section]
\newtheorem{corollary}[theorem]{Corollary}
\newtheorem{lemma}[theorem]{Lemma}
\newtheorem{proposition}[theorem]{Proposition}

\newproclaim{Remark}{Remark}

\newcommand{\ES}{\mathit{ES}}

\makeatother

\begin{document}
\begin{frontmatter}

\title{Correlation analysis of enzymatic reaction of a~single protein molecule\thanksref{T1}}
\runtitle{\hspace*{-5pt}Correlation analysis of single-molecule enzymatic reaction}

\thankstext{T1}{Supported in part by NSF Grant DMS-04-49204 and NIH/NIGMS Grant R01GM090202.}

\begin{aug}
\author[A]{\fnms{Chao} \snm{Du}}
\and
\author[A]{\fnms{S. C.} \snm{Kou}\corref{}\ead[label=e1]{kou@stat.harvard.edu}}
\runauthor{C. Du and S. C. Kou}
\affiliation{Harvard University}
\address[A]{Department of Statistics\\
Harvard University\\
Cambridge, Massachusetts 02138\\
USA\\
\printead{e1}} 
\end{aug}

\received{\smonth{4} \syear{2011}}
\revised{\smonth{1} \syear{2012}}

%
\begin{abstract}
New advances in nano sciences open the door for scientists to study
biological processes on a microscopic molecule-by-molecule basis. Recent
single-molecule biophysical experiments on enzyme systems, in particular,
reveal that enzyme molecules behave fundamentally differently from what
classical model predicts. A stochastic network model was previously proposed
to explain the experimental discovery. This paper conducts detailed
theoretical and data analyses of the stochastic network model, focusing on
the correlation structure of the successive reaction times of a single
enzyme molecule. We investigate the correlation of experimental fluorescence
intensity and the correlation of enzymatic reaction times, and examine the
role of substrate concentration in enzymatic reactions. Our study shows that
the stochastic network model is capable of explaining the experimental data
in depth.
\end{abstract}

%

\end{frontmatter}

\section{Introduction}

In a chemical reaction, the number of molecules involved can
drastically vary from millions of moles---a forest devastated by a
fire---to only a few---reactions in a living cell. While most
conventional chemical experiments were designed for a large ensemble in
which only the average could be observed, chemistry textbooks tend to
explain what really happens in a reaction on a molecule-by-molecule
basis. This extrapolation certainly requires the homogeneity
assumption: each molecule behaves in the same way, so the average also
represents individual behavior. To verify this assumption, the kinetic
of a single molecule must be directly observed, which requires rather
sophisticated technology not available until the 1990s. Since then, the
development of nanotechnology has enabled scientists to track and
manipulate molecules\vadjust{\goodbreak} one by one. A new age of single-molecule
experiments began [\citet{NieZar97}, \citet{XieTra98},
\citet{XieLu99}, \citet{Tametal}, \citet{Wei00},
\citet{Moe}, \citet{Flo05}, \citet{KouXieLiu05},
\citet{Kou09}].

Such experiments offer a greatly amplified view of single-molecular
dynamics over considerably \textit{long} time periods from seconds to
hours, a time scale that far exceeds what can be achieved by computer
based molecular dynamic simulation (even with a super computer,
molecular dynamic simulation cannot reach beyond milliseconds). The
single-molecule experiments also provide detailed information on the
intermediate transition steps of a~biological process not available in
traditional experiments. Not surprisingly, these experiments reveal the
stochastic nature of nanoscale particles long masked by ensemble
averages: rather than remain rigid, those particles undergo dramatic
conformation change driven by external thermal motion. Future
development in this area will provide us a deeper understanding of
biological processes [such as molecular motors,
\citet{AsbFehBlo03}] and accelerate new technology development
[such as single-molecule gene sequencing, \citet{PusNefQua09}].

Among bio-molecules, enzymes play an important role: by lowering the energy
barrier between the reactant and product, they ensure that many life
essential processes can be effectively carried out in a living cell. An
aspiration of bioengineers is to artificially design and produce new and
efficient enzymes for specific use. Studying and understanding the mechanism
of existing enzymes, therefore, remains one of the central topics in life
science. According to the classical literature, the kinetic of an
enzyme is
described by the Michaelis--Menten mechanism [\citet{AtkdeP02}]: an
enzyme molecule $E$ could bind with a reactant molecule $S$, which is
referred to as a substrate in the chemistry literature (hence the
symbol $S$%
), to form a complex $\ES$. The complex can either dissociate to enzyme
and substrate molecules or undergo a catalytic process to release the
product~$P$. The enzyme then returns to the original state $E$ to start
another
catalytic circle. This process is typically diagrammed as%
%
%
\begin{equation} \label{MM}
E+S\underset{k_{-1}}{\stackrel{k_{1}[S]}{\rightleftarrows}}\ES\stackrel
{k_{2}}{%
\rightarrow}E^{0}+P,\qquad E^{0}\stackrel{\delta}{\rightarrow}E,
\end{equation}
where $[S]$ is the substrate concentration ($E^{0}$ is the release
state of
the enzyme), $k_{1}$~is the association rate per unit substrate
concentration, $k_{-1}$ and~$k_{2}$ are, respectively, the dissociation and
catalytic rate, and $\delta$ is the returning rate. All the
transitions are
memoryless in the Michaelis--Menten scheme, so the whole process can be
modeled as a continuous-time Markov chain consisting of three states
$E$, $%
\ES $ and $E^{0}$ for an enzyme molecule.

\begin{figure}

\includegraphics{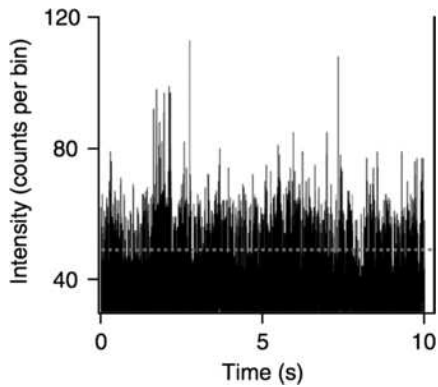}

\caption{Fluorescence intensity reading from one experiment (the substrate
concentration is 100 micro-molar). Each fluorescence intensity spike is
caused by the release of a reaction product.}
\label{fig1}
\end{figure}

A recent single-molecule experiment [\citet{Engetal06}] conducted
by the Xie group at Harvard University (Department of Chemistry and Chemical
Biology) studied the enzyme $\beta$-galactosidase ($\beta$-gal), which
catalyzes the breakdown\vadjust{\goodbreak} of the sugar lactose and is essential in the human
body [\citet{Jacetal94}, \citet{Dor03}]. In the experiment a
single $\beta$-gal molecule is immobilized (by linking to a bead bound
on a
glass coverslip) and immersed in buffer solution of the substrate molecules.
This setup allows $\beta$-gal's enzymatic action to be continuously
monitored under a fluorescence microscope. To detect the individual
turnovers, that is, the enzyme's switching from the~$E$ state to the~$E^{0}$
state, careful design and special treatment were carried out (such as the
use of photogenic substrate resorun-$\beta$-D-galactopyranoside) so that
once the experimental system was placed under a laser beam the reaction
product and \textit{only} the reaction product was fluorescent. This setting
ensures that as the $\beta$-gal enzyme catalyzes substrate molecules one
after another, a strong fluorescence signal is emitted and detected
only when
a~product is released, that is, only when the reaction reaches the $E^{0}+P$
stage in~(\ref{MM}). Recording the fluorescence intensity over time thus
enables the experimental determination of individual turnovers. A sample
fluorescence intensity trajectory from this experiment is shown in
Figure %
\ref{fig1}. High spikes in the trajectory are the results of intense photon
burst at the $E^{0}+P$ state, while low readings correspond to the $E$
or $%
\ES $ state. The time lag between two adjacent high fluorescence spikes is
the enzymatic turnover time, that is, the time to complete a catalytic circle.

%
%

%
%
\begin{figure}

\includegraphics{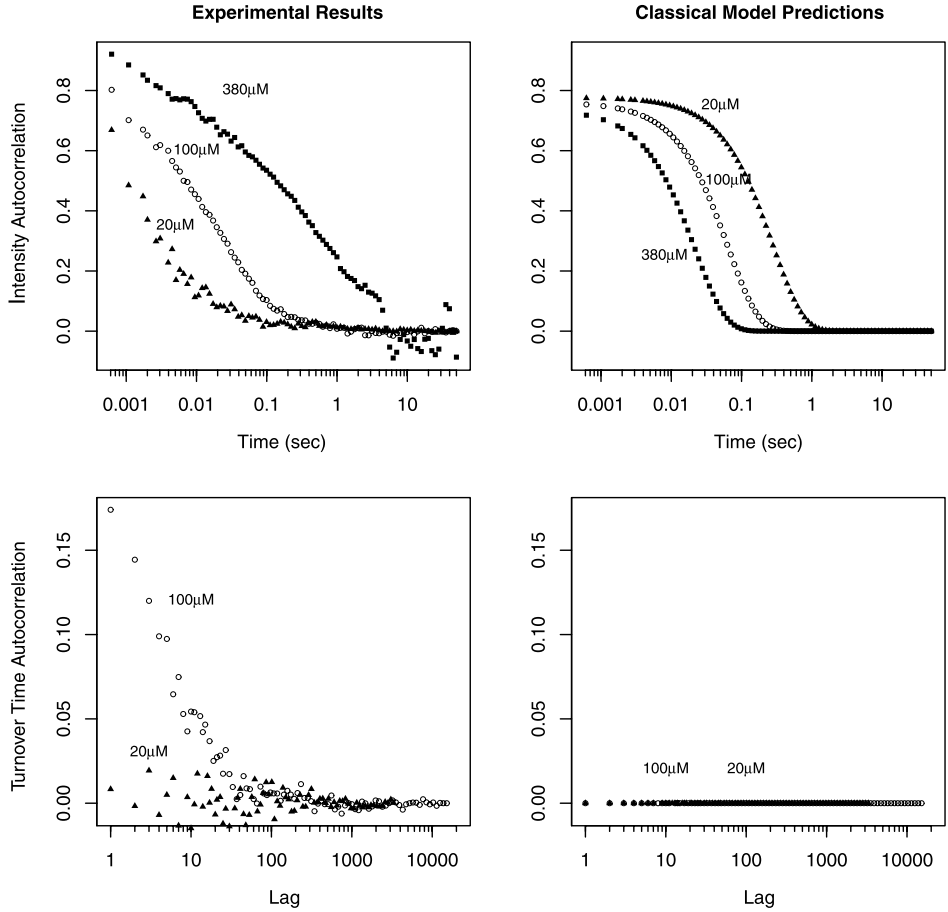}

\caption{Left column: experimentally observed fluorescence intensity and
turnover time autocorrelations under different substrate concentrations $[S]$
(20, 100 and 380 micromolar). Right column: the autocorrelations predicted
by the classical Michaelis--Menten model. Under the Michaelis--Menten model,
the turnover time autocorrelation should be zero and the intensity
autocorrelations should decay exponentially and decay faster under larger
concentration. All contradict the experimental findings.}
\label{fig2}
\end{figure}

Examining the experimental data, including the distribution and
autocorrelation of the turnover times as well as the fluorescence intensity
autocorrelation, researchers were surprised that the experimental data
showed a considerable departure from the Michaelis--Menten mechanism.
Section %
\ref{sec2} describes the experimental findings in detail. Figure~\ref{fig2}
illustrates the discrepancy between the experimental data and the
Michaelis--Menten model in terms of the autocorrelations. The left two panels
show the experimentally observed fluorescence intensity autocorrelation and
turnover time\vadjust{\goodbreak} autocorrelation under different substrate concentrations
$[S]$. The right two panels show the corresponding autocorrelation patterns
predicted by the Michaelis--Menten model. Comparing the bottom two
panels, we
note that under the classical Michaelis--Menten model the turnover time
autocorrelation should be zero (hence the horizontal line at the
bottom-right panel), which clearly contradicts the experimental result on
the left. From the top two panels we note that under the Michaelis--Menten
model the fluorescence intensity autocorrelation should decay \textit{%
exponentially} and should decay \textit{faster} with \textit{larger} substrate
concentration, but the experimental result shows the opposite: the
intensity autocorrelations decay \textit{slower} with \textit{larger} substrate
concentration, and they do not decay exponentially.

To explain the experimental puzzle, a new stochastic network model was
introduced [\citet{Kouetal}, \citet{Kou08N2}], and it was
shown that the
stochastic network model well explained the experimental distribution
of the
turnover times. The autocorrelation of successive turnover times and the
correlation of experimental fluorescence intensity, however, were not
investigated in the previous articles.

This paper further explores the stochastic network model, concentrating on
the correlation structure of the turnover times and that of the fluorescence
intensity. The rest of the paper is organized as follows. Section~\ref{sec2}
reviews the preceding work, including the experiment observation and
the new
stochastic network model. Section~\ref{sec3} analytically calculates the
turnover time autocorrelation and the fluorescence intensity autocorrelation
based on the stochastic network model. These analytical results give an
explanation of the multi-exponentially decay pattern of the autocorrelation
functions. Section~\ref{sec4} discusses how to fit the experiment data
within the framework of the stochastic network model. The paper ends in
Section~\ref{sec5} with a summary and some concluding remarks.

\section{Modeling enzymatic reaction}
\label{sec2}

\subsection{The classical model and its challenge}

Under the classical Michaelis--Menten model~(\ref{MM}), an enzyme molecule
behaves as a three-state continuous-time Markov chain with the generating
matrix (infinitesimal generator)%
\[
\mathbf{Q}_{\mathrm{MM}}=
\pmatrix{
-k_{1}[S] & k_{1}[S] & 0 \cr
k_{-1} & -(k_{-1}+k_{2}) & k_{2} \cr
\delta& 0 & -\delta}.
\]
We can readily draw two properties from this continuous-time Markov chain
model.
%
%
\begin{proposition}
The density function of the turnover time, the time that it takes the enzyme
to complete one catalytic cycle (i.e., to go from state~$E$ to state~$E^{0}$%
), is%
\[
f(t)=\frac{k_{1}k_{2}[S]}{2p}\bigl(e^{-(q-p)t}-e^{-(q+p)t}\bigr),
\]
where $p=\sqrt{(k_{1}[S]+k_{2}+k_{-1})^{2}/4-k_{1}k_{2}[S]}$ and $%
q=(k_{1}[S]+k_{2}+k_{-1})/2$.
\end{proposition}
%
%
\begin{proposition}
The successive turnover times have no correlation.
\end{proposition}

The first proposition implies that the density of turnover time is
almost an
exponential, since the term $e^{-(q-p)t}$ easily dominates the\vadjust{\goodbreak} term $%
e^{-(q+p)t}$ for most values of $t$; see \citet{Kou08N2} for a
proof. The second
proposition is a consequence of the Markov property: each turnover time,
which is a first passage time, is independently and identically distributed.

The third property concerns the autocorrelation of the fluorescence
intensity. As we have seen in Figure~\ref{fig1}, the experimentally recorded
fluorescence intensity consists of high spikes and low readings. The high
peaks correspond to the release of the fluorescent product (when the enzyme
is at the state $E^{0}$), whereas the low readings come from the background
noise. We can thus think of the fluorescence intensity reading as a record
of an on--off system: $E^{0}$ being the on state, $E$ and $\ES$ being the off
states.
%
%
\begin{proposition}
The autocorrelation function of the fluorescence intensity is proportional
to $\exp(-t(k_{-1}+k_{2}+k_{1}[S]))$.
\end{proposition}

The proof of the proposition will be given in Corollary~\ref{cor4}. This
proposition says that under the Michaelis--Menten model the intensity
autocorrelation decays exponentially and faster with larger substrate
concentration~$[S]$.

The results from the single-molecule experiment on $\beta$-gal
[\citet{Engetal06}] contradict all three properties of the
Michaelis--Menten model:\vspace*{8pt}

(1) The empirical distribution of the turnover time does not exhibit
exponential decay; see \citet{Kou08N2} for a detailed explanation.

(2) The experimental turnover time autocorrelations are far from zero, as
seen in Figure~\ref{fig2}.

(3) The experimental intensity autocorrelations decay neither exponentially
nor faster under larger concentration. See Figure~\ref{fig2}.

\subsection{A stochastic network model}

We believe these contradictions are rooted in the molecule's
\textit{dynamic} conformational fluctuation. An enzyme molecule is not
rigid: it experiences constant changes and fluctuations in its
three-dimensional shape and configuration due to the entropic and
atomic forces at the nano scale [\citet{KouXie04},
\citet{Kou08N1}]. Although for a large ensemble of molecules, the
(nanoscale) conformational fluctuation is buried in the macroscopic
population average, for a single molecule the conformational
fluctuation can be much more pronounced: different conformations could
have different chemical properties, resulting in time-varying
performance of the enzyme, which can be studied in the single-molecule
experiment. The following stochastic network model
[\citet{Kouetal}] was developed with this
idea:%
%
%
\begin{eqnarray} \label{2by2}
&&S+E_{1} \underset{k_{-11}}{\stackrel{k_{11}[S]}{\rightleftarrows}} \ES_{1}
\stackrel{k_{21}}{\rightarrow} P+E_{1}^{0},\qquad E_{1}^{0}%
\stackrel{\delta_{1}}{\rightarrow}E_{1}, \nonumber\\
&&\hphantom{S}\downarrow\uparrow\hspace*{37pt}\downarrow\uparrow\hspace*{30pt}\hphantom{0}
\downarrow\uparrow\qquad\hspace*{10pt}\hphantom{0}\cdots\nonumber\\
&&S+E_{2} \underset{k_{-12}}{\stackrel{k_{12}[S]}{\rightleftarrows}} \ES_{2}
\stackrel{k_{22}}{\rightarrow} P+E_{2}^{0},\qquad E_{2}^{0}\stackrel
{\delta_{2}}{\rightarrow}E_{2},\\
&&\hphantom{S}\hspace*{1.5pt}\cdots\hspace*{37pt}\cdots\hspace*{33.5pt}\hphantom{0}
\cdots\qquad\hspace*{10.5pt}\hphantom{0}\cdots\nonumber\\[-8pt]
&&\hphantom{S}\downarrow\uparrow\hspace*{37pt}\downarrow\uparrow\hspace*{30pt}\hphantom{0}
\downarrow\uparrow\qquad\hspace*{10pt}\hphantom{0}\cdots\nonumber\\
&&S+E_{n} \underset{k_{-1n}}{\stackrel{k_{1n}[S]}{\rightleftarrows}} \ES_{n}
\stackrel{k_{2n}}{\rightarrow} P+E_{n}^{0},\qquad E_{n}^{0}\stackrel
{\delta_{n}}{\rightarrow}E_{n}.\nonumber
\end{eqnarray}
This is still a Markov chain model but with $3n$ states instead of three.
The enzyme still exists as a free enzyme $E$, an enzyme--substrate
complex $%
\ES $ or a returning enzyme $E^{0}$, but it can take $n$ different
conformations indexed by subscripts in each stage. At each transition, the
enzyme can either change its conformation within the same stage (such
as $%
E_{i}\rightarrow E_{j}$ or $\ES_{i}\rightarrow \ES_{j}$) or carry out one
chemical step, that is, move between the stages (such as
$E_{i}\rightarrow
\ES_{i}$, $\ES_{i}\rightarrow E_{i}$ or $\ES_{i}\rightarrow E_{i}^{0}$). Since
only the product $P$ is fluorescent in the experiment, in model~(\ref{2by2})
any state $E_{i}^{0}$ is an on-state, and the others are off-states.
Consequently, the turnover time is the traverse time between any two
on-states $E_{j}^{0}$ and $E_{k}^{0}$.

To fully specify the model, we need to stipulate the transition rates.
For $%
i\neq j$, we use $\alpha_{ij}$, $\beta_{ij}$ and $\gamma_{ij}$ to denote,
respectively, the transition rates of $E_{i}\rightarrow E_{j}$, $%
\ES_{i}\rightarrow \ES_{j}$ and $E_{i}^{0}\rightarrow E_{j}^{0}$.
$k_{1i}[S]$%
, $k_{-1i}$, $k_{2i}$ and $\delta_{i}$ are, respectively, the transition
rates of $E_{i}\rightarrow \ES_{i}$, $\ES_{i}\rightarrow E_{i}$, $%
\ES_{i}\rightarrow E_{i}^{0}$ and $E_{i}^{0}\rightarrow E_{i}$. Define $%
\mathbf{Q}_{AA}$, $\mathbf{Q}_{BB}$ and $\mathbf{Q}_{CC}$ to be square
matrices:%
\[
\mathbf{Q}_{AA}=[\alpha_{ij}]_{n\times n},\qquad \mathbf{Q}_{BB}=[\beta
_{ij}]_{n\times n},\qquad \mathbf{Q}_{CC}=[\gamma_{ij}]_{n\times n},
\]
where $\alpha_{ii}=-\sum_{j\neq i}\alpha_{ij}$, $\beta_{ii}=-\sum
_{j\neq
i}\beta_{ij}$ and $\gamma_{ii}=-\sum_{j\neq i}\gamma_{ij}$. They
correspond to transitions among the $E_{i}$ states, among the $\ES_{i}$
states and among the $E_{i}^{0}$ states, respectively. Define diagonal
matrices%
%
%
\begin{eqnarray} \label{Qsub}\qquad
\mathbf{Q}_{AB} &=&\operatorname{diag}\{k_{11}[S],k_{12}[S],\ldots
,k_{1n}[S]\},\nonumber\\
\mathbf{Q}_{BA}&=&\operatorname{diag}\{
k_{-11},k_{-12},\ldots,k_{-1n}\},\nonumber\\[-8pt]\\[-8pt]
\mathbf{Q}_{BC} &=&\operatorname{diag}\{k_{21},k_{22},\ldots,k_{2n}\},\nonumber\\
\mathbf{Q}_{CA}&=&\operatorname{diag}\{\delta_{1},\delta_{2},\ldots,\delta
_{n}\}.
\nonumber
\end{eqnarray}
They correspond to transitions between the different stages. The generating
matrix of model~(\ref{2by2}) is then%
%
%
\begin{equation}\label{Q}\quad
\mathbf{Q}=\pmatrix{
\mathbf{Q}_{AA}-\mathbf{Q}_{AB} & \mathbf{Q}_{AB} & \mathbf{0} \cr
\mathbf{Q}_{BA} & \mathbf{Q}_{BB}-(\mathbf{Q}_{BA}+\mathbf{Q}_{BC}) &
\mathbf{Q}_{BC} \cr
\mathbf{Q}_{CA} & \mathbf{0} & \mathbf{Q}_{CC}-\mathbf{Q}_{CA}}.
\end{equation}

Under this new model, the distribution of the turnover time, the correlation
of turnover times and the correlation of the fluorescence intensity can be\vadjust{\goodbreak}
analyzed and compared with experimental data. This paper studies the
autocorrelation of turnover time and the autocorrelation of fluorescence
intensity.

\section{Autocorrelation of turnover time and of fluorescence intensity}
\label{sec3}

\subsection{Dynamic equilibrium and stationary distribution}

In the chemistry literature, the term ``equilibrium'' often refers to
the state in which all the macroscopic quantities of a system are
time-independent. For the microscopic system studied in single-molecule
experiments, macroscopic quantities, however, are meaningless, and
microscopic parameters never cease to fluctuate. Nonetheless, for a
micro-system, one can talk about dynamic equilibrium in the sense that
the \textit{distribution} of the state quantities become
time-independent, that is, they reach the stationary distribution. The
single-molecule enzyme experiment that we consider here falls into this
category, since the enzymatic reactions happen quite fast. We cite the
following lemma [Lemma 3.1 of \citet{Kou08N2}], which gives the
stationary distribution of the Markov chain~(\ref{2by2}):
%
%
\begin{lemma}
\label{lem1}Let $X(t)$ be the process evolving according to~(\ref{2by2}).
Suppose all the parameters $k_{1i},k_{-1i},k_{2i},\delta_{i},\alpha
_{ij},\beta_{ij}$, and $\gamma_{ij}$ are positive. Then $X(t)$ is ergodic.
Let the row vectors $\bolds{\pi}_{A}=(\pi(E_{1}),\pi(E_{2}),\ldots,\pi
(E_{n}))$, $\bolds{\pi}_{B}=(\pi(\ES_{1}),\ldots,\pi(\ES_{n}))$, and
$\bolds
{\pi}%
_{C}=(\pi(E_{1}^{0}),\ldots,\pi(E_{n}^{0}))$ denote the stationary
distribution of the entire network. Up to a normalizing constant, they are
determined by%
\begin{eqnarray*}
&\displaystyle \bolds{\pi}_{A}=-\bolds{\pi}_{C}\mathbf{Q}_{CA}\mathbf{L},\qquad \bolds{\pi
}_{B}=-%
\bolds{\pi}_{C}\mathbf{Q}_{CA}\mathbf{M},& \\
&\displaystyle \bolds{\pi}_{C}(\mathbf{Q}_{CC}-\mathbf{Q}_{CA}-\mathbf{Q}_{CA}\mathbf
{MQ}%
_{BC})=0,&
\end{eqnarray*}
where the matrices%
\begin{eqnarray*}
\mathbf{L} &=&[ \mathbf{Q}_{AA}-\mathbf{Q}_{AB}-\mathbf{Q}_{AB}(\mathbf{
Q}_{BB}-\mathbf{Q}_{BA}-\mathbf{Q}_{BC})^{-1}\mathbf{Q}_{BA}] ^{-1}, \\
\mathbf{M} &=&[ \mathbf{Q}_{BB}-\mathbf{Q}_{BC}-(\mathbf{Q}_{BB}-%
\mathbf{Q}_{BA}-\mathbf{Q}_{BC})\mathbf{Q}_{AB}^{-1}\mathbf{Q}_{AA}]
^{-1}.
\end{eqnarray*}
\end{lemma}

Under the stochastic network model~(\ref{2by2}), a turnover event can start
from any state $E_{i}$ and end in any $E_{j}^{0}$. It follows that the
\textit{%
overall} distribution of all the turnover times is characterized by a
mixture distribution with the weights given by the \textit{stationary}
probability of a turnover event's starting from $E_{i}$. The following
lemma, based on Lemma 3.4 of \citet{Kou08N2}, provides the stationary
probability.
%
%
\begin{lemma}
\label{lem2}Let $\mathbf{w}$ be a row vector, $\mathbf{w}%
=(w(E_{1}),w(E_{2}),\ldots, w(E_{n}))$, where $w(E_{i})$ denotes the
stationary probability of a turnover event's starting from state~$E_{i}$.
Then up to a normalizing constant, $\mathbf{w}$ is the nonzero solution
of%
%
%
\begin{equation} \label{eqV}
\mathbf{w}(\mathbf{I}+\mathbf{MQ}_{BC}-\mathbf{Q}_{CA}^{-1}\mathbf{Q}%
_{CC})=0.\vadjust{\goodbreak}
\end{equation}
\end{lemma}

\subsection{Autocorrelation of turnover time}

\subsubsection*{Expectation of turnover time}

The enzyme turnover event occurs one after another. Each can start from
any $%
E_{i}$ and end in any $E_{j}^{0}$. The next turnover may start from
$E_{k}$ (%
$k\neq j$) when the system exits the $E^{0}$ stage from~$E_{k}^{0}$. To
calculate the correlation between turnover times, it is necessary to find
out the probabilities of all these combinations and the expected turnover
times. We introduce the following notation.

Let $T_{E_{i}}$ and $T_{\ES_{i}}$ denote the first passage time of reaching
the set $\{E_{1}^{0},\allowbreak E_{2}^{0},\ldots, E_{n}^{0}\}$ from $E_{i}$ and
$\ES_{i}$%
, respectively. Let $P_{E_{i}E_{j}^{0}}$ and $P_{\ES_{i}E_{j}^{0}}$ be the
probability\vspace*{-1pt} that a turnover event, starting, respectively, from $E_{i}$
and $%
\ES_{i}$, ends in $E_{j}^{0}$. Let $P_{E_{i}^{0}E_{j}}$ denote the
probability that, after the previous turnover ends in~$E_{i}^{0}$, a new
turnover event starts from $E_{j}$. Finally, let $T_{E_{i}E_{j}^{0}}$
and $%
T_{\ES_{i}E_{j}^{0}}$ be the first passage time of reaching the state $%
E_{j}^{0}$ from $E_{i}$ and $\ES_{i}$, respectively.\vspace*{1pt}

For the values of $E(T_{E_{i}})$ and $E(T_{\ES_{i}})$, we cite the following
lemma [Corollary~3.3 of \citet{Kou08N2}].
%
%
\begin{lemma}
\label{lem3}Let the vectors $\bolds{\mu
}_{A}=(E(T_{E_{1}}),E(T_{E_{2}}),\ldots
,E(T_{E_{n}}))^{T}$ and $\bolds{\mu}_{B}=(E(T_{\ES_{1}}),\ldots
,E(T_{\ES_{n}}))^{T}$ denote the mean first passage times. Then they are
given by%
%
%
\begin{equation}\label{meanvector}
\pmatrix{
\bolds{\mu}_{A} \cr
\bolds{\mu}_{B}}
=\pmatrix{
-(\mathbf{L}+\mathbf{M})\mathbf{1} \cr
-(\mathbf{N}+\mathbf{R})\mathbf{1}},
\end{equation}
where the matrices $\mathbf{N}$ and $\mathbf{R}$ are given by%
\begin{eqnarray*}
\mathbf{N} &=&[ \mathbf{Q}_{AA}-(\mathbf{Q}_{AA}-\mathbf{Q}_{AB})%
\mathbf{Q}_{BA}^{-1}(\mathbf{Q}_{BB}-\mathbf{Q}_{BC})] ^{-1}, \\
\mathbf{R} &=&[ \mathbf{Q}_{BB}-\mathbf{Q}_{BA}-\mathbf{Q}_{BC}-\mathbf{
Q}_{BA}(\mathbf{Q}_{AA}-\mathbf{Q}_{AB})^{-1}\mathbf{Q}_{AB}] ^{-1}.
\end{eqnarray*}
\end{lemma}

For the probabilities $P_{E_{i}E_{j}^{0}}$, $P_{\ES_{i}E_{j}^{0}}$ and $%
P_{E_{i}^{0}E_{j}}$, we have the following lemma:
%
%
\begin{lemma}
\label{lem4}Let $\mathbf{P}_{AC}$, $\mathbf{P}_{BC}$ and $\mathbf{P}_{CA}$
be probability matrices $\mathbf{P}_{AC}=[P_{E_{i}E_{j}^{0}}]_{n\times
n}$, $%
\mathbf{P}_{BC}=[P_{\ES_{i}E_{j}^{0}}]_{n\times n}$ and $\mathbf{P}%
_{CA}=[P_{E_{i}^{0}E_{j}}]_{n\times n}$. Then they are given by%
%
%
\begin{equation} \label{pmat}
\mathbf{P}_{AC}=-\mathbf{MQ}_{BC},\qquad \mathbf{P}_{BC}=-\mathbf{RQ}%
_{BC},\qquad \mathbf{P}_{CA}=(\mathbf{I}-\mathbf{Q}_{CA}^{-1}\mathbf{Q}%
_{CC})^{-1}.\hspace*{-35pt}
\end{equation}
\end{lemma}

For the expectation of $T_{E_{i}E_{j}^{0}}$ and $T_{\ES_{i}E_{j}^{0}}$, we
have the following.
%
%
\begin{lemma}
\label{lem5}Let
\[
\mathbf{E}%
_{AC}=[P_{E_{i}E_{j}^{0}}E(T_{E_{i}E_{j}^{0}})]_{n\times n}\quad \mbox{and}
\quad\mathbf{E}%
_{BC}=[P_{\ES_{i}E_{j}^{0}}E(T_{\ES_{i}E_{j}^{0}})]_{n\times n}
\]
be two $n\times n$ matrices. Then they are given by%
\[
\mathbf{E}_{AC}=(\mathbf{LM}+\mathbf{MR})\mathbf{Q}_{BC},\qquad \mathbf{E}%
_{BC}=(\mathbf{NM}+\mathbf{RR})\mathbf{Q}_{BC}.\vadjust{\goodbreak}
\]
\end{lemma}

We defer the proofs of Lemmas~\ref{lem4} and~\ref{lem5} to the \hyperref
[app]{Appendix}.

\subsubsection*{Correlation of the turnover times}

Let $T^{i}$ denote the $i$th turnover time. The next theorem, based on
Lemmas~\ref{lem1} to~\ref{lem5}, obtains the autocorrelation of the
successive turnover times. We defer its proof to the \hyperref[app]{Appendix}.
%
%
\begin{theorem}
\label{theo1}The covariance between the first turnover and the $m$th
turnover ($m>1$) is given by%
\[
\operatorname{cov}(T^{1},T^{m})=-\mathbf{w}\bigl(\mathbf{L}+\mathbf
{M}(\mathbf{I}-%
\mathbf{Q}_{AB}^{-1}\mathbf{Q}_{AA})\bigr)[(\mathbf{P}_{AC}\mathbf
{P}_{CA})^{m-1}-%
\mathbf{1w}]\bolds{\mu}_{A},
\]
where $\mathbf{P}_{AC}\mathbf{P}_{CA}=-\mathbf{MQ}_{BC}(\mathbf
{I}-\mathbf{Q}%
_{CA}^{-1}\mathbf{Q}_{CC})^{-1}$.
\end{theorem}

The matrix $\mathbf{P}_{AC}\mathbf{P}_{CA}$ is the product of two
transition-probability matrices, so it is a stochastic matrix. Given that
all the states in the stochastic network model communicate with each
other, $%
\mathbf{P}_{AC}\mathbf{P}_{CA}$ is also irreducible, and all its
entries are
positive. According to the Perron--Frobenius theorem [\citet{HorJoh85}],
such a matrix has eigenvalue one with simplicity one, and the absolute values
of the other eigenvalues are strictly less than one. We therefore
obtain the
following corollary of Theorem~\ref{theo1}.
%
%
\begin{corollary}
\label{cor1}Suppose that $\mathbf{P}_{AC}\mathbf{P}_{CA}$ is diagonalizable:
\[
\mathbf{P}_{AC}\mathbf{P}_{CA}=\mathbf{U}\bolds{\lambda}\mathbf
{U}^{-1}=\mathbf{%
1w}+\sum_{l=2}^{n}\lambda_{l}\varphi_{l}\psi_{l}^{T},
\]
where the diagonal matrix $\bolds{\lambda}=\operatorname{diag}(1,\lambda_{2},\ldots
,\lambda
_{n})$ consists of the eigenvalues of $\mathbf{P}_{AC}\mathbf{P}_{CA}$
with $%
|\lambda_{i}|<1$; the columns, $\mathbf{1},\varphi_{2},\ldots,\varphi
_{n} $, of matrix $\mathbf{U}$ are the corresponding right
eigenvectors; and
the rows, $\mathbf{w},\psi_{2},\ldots,\psi_{n}$, of $\mathbf{U}^{-1}$ are
the corresponding left eigenvectors. Then we have%
%
%
\begin{equation}\label{corrT}
\operatorname{cov}(T^{1},T^{m})=\sum_{i=2}^{n}\sigma_{i}\lambda_{i}^{m-1},
\end{equation}
where $\sigma_{i}=-\mathbf{w}(\mathbf{L}+\mathbf{M}(\mathbf{I}-\mathbf
{Q}%
_{AB}^{-1}\mathbf{Q}_{AA}))\varphi_{i}\psi_{i}^{T}\bolds{\mu}_{A}$.
\end{corollary}

Although the matrix $\mathbf{P}_{AC}\mathbf{P}_{CA}$ may have complex
eigenvalues, these complex eigenvalues and corresponding eigenvectors always
appear as conjugate pairs so that the imaginary parts in (\ref%
{corrT}) cancel each other. As a result, we could treat all $\lambda_{i}$
and $\sigma_{i}$ as if they were real numbers.

Theorem~\ref{theo1}, along with Corollary~\ref{cor1}, provides an
explanation of why the correlation of turnover times is not zero. At first
sight, it seems to contradict the memoryless property of a Markov chain.
What actually happens is that the\vadjust{\goodbreak} state must be explicitly specified
for the
memoryless property to hold (i.e., one needs to exactly specify whether an
enzyme is at state $E_{1}$ or $E_{2}$), whereas in the single-molecule
experiment we only know whether the system is in an ``on'' or ``off''
state (e.g., one only knows that the enzyme is in one of the on-states $%
E_{1}^{0},\ldots, E_{n}^{0}$). When there are multiple states, this
aggregation effect leads to incomplete information that prevents the
independence between successive turnovers; consequently, each turnover time
carries some information about its reaction path, which is correlated with
the reaction path of the next turnover, resulting in the correlation between
successive turnover times.

Corollary~\ref{cor1} also states that since $|\lambda_{i}|<1$, the
autocorrelation is a mixture of exponential decays. Thus, depending on the
relative scales of the eigenvalues, the actual decay might be
single-exponential when one eigenvalue dominates the others or
multi-exponential when several major eigenvalues jointly contribute to the
decay.\vspace*{-2pt}

\subsubsection*{Fast enzyme reset}

In most enzymatic reactions, including the one we study, the enzyme
returns very quickly to restart a new cycle once the product is
released [\citet{Seg75}]. Those enzymes are called
fast-cycle-reset enzymes. To model this
fact, we let $\delta_{i}$ $(i=1,2,\ldots,n)$, the transition rate
from~$E_{i}^{0}$ to $E_{i}$, go to infinity. Then any enzyme in state $E_{i}^{0}$
will always return to state $E_{i}$ instantly, and the related transition
probability matrix~$\mathbf{P}_{CA}$, defined in~(\ref{pmat}),
becomes the identity matrix.\vspace*{-2pt}

\subsection{Autocorrelation of fluorescence intensity}\vspace*{-2pt}
\subsubsection*{Correlation of intensity as a function of time}

In the single-enzyme experiments, the raw data are the time traces of
fluorescence intensity, as shown in Figure~\ref{fig1}. The time lag between
two adjacent high fluorescence spikes gives the enzymatic turnover
time. The
fluorescence intensity reading, however, is subject to detection error: the
error caused by the limited time resolution $\Delta t$ of the detector.
Starting from time 0, the detector will only record intensity data at
multiples of~$\Delta t$: $0,\Delta t,2\Delta t,\ldots,k\Delta
_{t},\ldots.$ The intensity reading at time $k\Delta t$ is actually the total
number of
photons received during the period of $((k-1)\Delta t,k\Delta t)$.
Thus, the
detection errors of turnover time are roughly $\Delta t$. When the successive
reactions occur slowly, the average turnover time is much longer than $%
\Delta t$, and the error is negligible. But when the reactions happen very
frequently, the average turnover time becomes comparable to $\Delta t$, and
this error cannot be ignored. In fact, when the substrate concentration is
high enough, the enzyme will reach the ``on'' states so frequently that most
of the intensity readings are very high, making it impossible to reliably
determine the individual turnover times. Under this situation, it is
necessary to directly study the behavior of the raw intensity reading.

There are two main sources of the photons generated in the experiment: the
weak but perpetual\vadjust{\goodbreak} background noise and the strong but short-lived burst.
The number of photons received from two different sources can be
modeled as
two independent Poisson processes with different rates. We can use the
following equation to represent $I(t)$, the intensity recorded at time~$t$:%
%
%
\begin{equation} \label{I}
I(t)=N_{t}(T_{\mathrm{on}}(t))+N_{t}^{0}(\Delta t),
\end{equation}
where $N_{t}(s)$ and $N_{t}^{0}(s)$ represent the total number of photons
received due to the burst and background noise, respectively, within a
length $s$ subinterval of $(t-\Delta t,t)$; $T_{\mathrm{on}}(t)$ is the total time
that the enzyme system spends at the ``on'' states (any $E_{i}^{0}$) within
the time interval $(t-\Delta t,t)$. $N_{t}(s)$ and $N_{t}^{0}(s)$ are
independent Poisson processes with rates $\nu$ and $\nu_{0}$, respectively.
With this representation, we have the following theorem, whose proof is
deferred to the \hyperref[app]{Appendix}.
%
%
\begin{theorem}
\label{theo2}The covariance of the fluorescence intensity is
%
%
\begin{equation}\label{covI}
\operatorname{cov}(I(0),I(t))\propto\sum_{i=2}^{3n}C_{i}e^{\mu
_{i}(t-\Delta t)},
\end{equation}
where $\mu_{i}$ are the nonzero eigenvalues of the generating matrix $%
\mathbf{Q}$ defined in~(\ref{Q}), and $C_{i}$ are constants only
depending on $\mathbf{Q}$.
\end{theorem}

Since $-\mathbf{Q}$ is a semi-stable matrix [\citet{HorJoh85}], it
follows that the real parts of all $\mu_{k}$ $(k>1)$ are negative. For a
real matrix, the complex eigenvalues along with their eigenvectors always
appear in conjugate pairs; thus, the imaginary parts cancel each other
in (%
\ref{covI}) and only the real parts are left. Therefore, we know according
to Theorem~\ref{theo2} that the covariance of intensity will decay
multi-exponentially.

\subsubsection*{Fast enzyme reset and intensity autocorrelation}

A fast-cycle-reset enzyme jumps from state $E_{i}^{0}$ to $E_{i}$ with
little delay. A short burst of photons is released during the enzyme's short
stay at $E_{i}^{0}$. For fast-cycle-reset enzymes, the behavior of the whole
system can be well approximated by an alternative system, where only
states $%
E_{i}$ and $\ES_{i}$ $(i\in1,2,\ldots,n)$ exist: the transition rates among
the $E$'s, among the $\ES$'s, and from $E_{i}$ to $\ES_{i}$ are exactly the
same as in the original system, but the transition rate from $\ES_{i}$
to $%
E_{i}$ is changed from $k_{-1i}$ to $k_{-1i}+k_{2i}$, since once a
transition of $\ES_{i}\rightarrow E_{i}^{0}$ occurs, the enzyme quickly moves
to $E_{i}$. We can thus think of lumping $E_{i}^{0}$ and $E_{i}$
together to
form the alternative system, which has generating matrix%
%
%
\begin{equation} \label{K}
\mathbf{K}=\pmatrix{
\mathbf{Q}_{AA}-\mathbf{Q}_{AB} & \mathbf{Q}_{AB} \cr
\mathbf{Q}_{BA}+\mathbf{Q}_{BC} & \mathbf{Q}_{BB}
-(\mathbf{Q}_{BA}+\mathbf{Q}%
_{BC})} .
\end{equation}
$\mathbf{K}$ is also a negative semi-stable matrix with $2n$ eigenvalues,
one of which is zero. The following theorem details how well the eigenvalues
of $\mathbf{K}$ approximate those of $\mathbf{Q}$.
%
%
\begin{theorem}
\label{theo3}Assume $\mathbf{Q}_{CA}=\delta\operatorname{diag}\{
q_{1},\ldots
,q_{n}\}$, where $q_{1},\ldots,q_{n}$ are fixed constants, while
$\delta$
is large. Let $\kappa_{i}$ $(i=2,3,\ldots,2n)$ denote the nonzero
eigenvalues of $\mathbf{K}$, then for each $\kappa_{i}$, there exists an
eigenvalue $\mu_{i}$ of $\mathbf{Q}$ such that%
\[
|\mu_{i}-\kappa_{i}|=O(\delta^{-1/2}).
\]
The other $n$ eigenvalues of $\mathbf{Q}$ satisfy%
\[
|\mu_{i}+\delta q_{i-2n}|=O(1),\qquad i=2n+1,\ldots,3n.
\]
\end{theorem}

The proof is deferred to the \hyperref[app]{Appendix}. This theorem
says that for
fast-cycle-reset enzymes with large $\delta$, the first $2n-1$ nonzero
eigenvalues of $\mathbf{Q}$ can be approximated by the eigenvalues of $%
\mathbf{K}$, while the other $n$ eigenvalues $\mu_{2n+1},\ldots,\mu_{3n}$
of $\mathbf{Q}$ are of the same order of $\delta$. Since all the
eigenvalues have negative real parts, according to~(\ref{covI}),
the terms associated with $\mu_{2n+1},\ldots,\mu_{3n}$ decay much faster
so their contribution can be ignored. Thus, we have the following results
for the intensity autocorrelation.
%
%
\begin{corollary}
\label{cor3}For fast-cycle-reset enzymes ($\delta\rightarrow\infty$),
\[
\operatorname{cov}(I(0),I(t))\propto\sum_{i=2}^{2n}C_{i}e^{\kappa
_{i}(t-\Delta
t)},
\]
where $\kappa_{i}$ are the nonzero eigenvalues of matrix $\mathbf{K}$
defined in~(\ref{K}).
\end{corollary}
%
%
\begin{corollary}
\label{cor4}For the classic Michaelis--Menten model, where \mbox{$n=1$},%
\[
\mathbf{K}=\pmatrix{
-k_{1}[S] & k_{1}[S] \cr
k_{2}+k_{-1} & -k_{2}-k_{-1}}.
\]
The only nonzero eigenvalue is $-(k_{-1}+k_{2}+k_{1}[S])$. We thus have,
for fast-cycle-reset enzymes,%
\[
\operatorname{cov}(I(0),I(t))\propto
e^{-(k_{-1}+k_{2}+k_{1}[S])(t-\Delta t)}.
\]
\end{corollary}

\section{From theory to data}
\label{sec4}

We have shown in the preceding sections that the autocorrelation of turnover
times and the correlation of intensity follow%
\[
\operatorname{cov}(T^{1},T^{m})\propto\sum\lambda_{i}^{m-1}\sigma_{i},\qquad
\operatorname{cov}(I(0),T(t))\propto\sum e^{\kappa_{i}(t-\Delta t)}C_{i}.
\]

Before applying these equations to fit the experimental data, the following
problems must be addressed. First, we know so far that the decay patterns
must be multi-exponential, but we do not yet know how the eigenvalues are
related to the rate constants ($k_{1i}$, $k_{-1i}$, $k_{2i}$, etc.) and the
substrate concentration $[S]$, which is the only adjustable parameter
in the
experiment. Second, we do not know the expressions of the coefficients
($%
\sigma_{i}$ and $C_{i}$). Third, we do not know the number of distinct
conformations $n$. We only know that it must be large: each enzyme consists
of hundreds of vibrating atoms, and, as a whole, it expands and rotates in
the 3-dimensional space within the constraint of chemical bonds. We next
address these questions before fitting the experimental data.\vspace*{-2pt}

\subsection{Eigenvalues as functions of rate constants and substrate
concentration}
\label{sec41}

In the enzyme experiments, the transition rates are intrinsic
properties of
the enzyme and the enzyme--substrate complex; they are not subject to
experimental control. The only variable subject to experimental control is
the concentration of the substrate molecules $[S]$. The higher the
concentration, the more likely that the enzyme molecule could bind with a
substrate molecule to form a complex. This is why the association rate $
k_{1i}[S]$ (the rate of $E_{i}\rightarrow \ES_{i}$) is proportional to the
concentration. The experiments were repeated under different concentrations,
resulting in different decay patterns of the autocorrelation functions
as in
Figure~\ref{fig2}. A~successful theory should be able to explain the
relationship between concentration and autocorrelation decay pattern.

The concentration only affect the transition rates between $E_{i}$ and $
\ES_{i}$, which are denoted by $\mathbf{Q}_{AB}$ in~(\ref{Qsub}).
Define $\mathbf{\tilde{Q}}_{AB}=\operatorname{diag}\{
k_{1i},k_{2i},\ldots
,k_{ni}\} $, which is independent of $[S]$; then $\mathbf
{Q}_{AB}=[S]\mathbf{%
\tilde{Q}}_{AB}$.\vspace*{-2pt}

\subsubsection*{Four scenarios for simplication}

To delineate the relationship between~$[S]$ and the autocorrelation decay
pattern, we next simplify the generating matrices. Below are four scenarios
that we will consider. Each of the scenarios guarantees the classical
Michaelis--Menten equation, a hyperbolic relationship between the reaction
rate and the substrate concentration,%
%
%
\begin{equation} \label{hyper}
v=\frac{1}{E(T)}=\frac{1}{\mathbf{w}\bolds{\mu}_{A}}\propto\frac{[
S]}{%
[S]+C} \qquad\mbox{with some constant }C,
\end{equation}
which was observed in both the traditional and single-molecule enzyme
experiments [see \citet{Kouetal}, \citet{Engetal06} and
\citet{Kou08N2} for
detailed discussion]. Each scenario has its own biochemical implications.

\textit{Scenario} 1. There are no or negligible transitions among the $E_{i}$
states, that is, $\alpha_{ij}\rightarrow0$ for $i\neq j$.

\textit{Scenario} 2. There are no or negligible transitions among the $%
\ES_{i} $ states, that is, $\beta_{ij}\rightarrow0$ for $i\neq j$.

Scenarios 1 and 2 correspond to the so-called slow fluctuating enzymes
(whose conformation fluctuates slowly over time).

\textit{Scenario} 3. The transitions among the $E_{i}$ states are much
faster than the others, that is, $\mathbf{Q}_{AA}=\tau\mathbf{\tilde
{Q}}%
_{AA}$ and the scale $\tau\gg1$ is much larger than other transition rates.

\textit{Scenario} 4. The transitions among the $\ES_{i}$ states are much
faster than the others, that is, $\mathbf{Q}_{BB}=\tau\mathbf
{\tilde{Q}}%
_{BB}$ and the scale $\tau\gg1$ is much larger than other transition
rates.\vadjust{\goodbreak}

Scenarios 3 and 4 correspond to the so-called fast fluctuating enzymes
(whose conformations fluctuate fast).\vspace*{-2pt}
\begin{Remark*} In the previous work [\citet{Kou08N2}], there are
two other
scenarios, which can also give rise to the hyperbolic relationship (\ref
{hyper}): primitive enzymes, whose dissociate rate is much larger than
their catalytic rate [\citet{AlbKno76}, \citet{Minetal},
\citet{Minetal05N1}], and conformational-equilibrium enzymes,
whose energy-barrier difference between dissociation and catalysis is
invariant across conformations [\citet{Minetal}]. But our analysis
based on those two scenarios does not lead to any meaningful
conclusion, so we omit them here.\vspace*{-2pt}
\end{Remark*}

\subsubsection*{The effect of concentration on turnover time autocorrelation}

Based on the four scenarios, we have the following theorem for
autocorrelation of turnover times.\vspace*{-2pt}
%
%
\begin{theorem}
\label{theo4}For enzymes with fast cycle reset, the transition probability
matrix governing the autocorrelation of turnover times is%
\[
\mathbf{P}_{AC}\mathbf{P}_{CA}=-\mathbf{MQ}_{BC}(\mathbf{I}-\mathbf{Q}%
_{CA}^{-1}\mathbf{Q}_{CC})^{-1}=-\mathbf{MQ}_{BC}.
\]
Its eigenvalues $\lambda_{i}$, under the four different scenarios,
satisfy the following:

\textup{Scenario} 1. $\lambda_{i}$ do not depend on $[S]$, the substrate
concentration. Thus, the autocorrelation decay should be similar for all
concentrations.

\textup{Scenario} 2. $\lambda_{i}$ depend on $[S]$ hyperbolically. More
precisely, if we use $\lambda_{i}([S])$ $(i=1,2,\ldots,n)$ to emphasize
the dependence of the eigenvalues on~$[S]$, we have%
\[
\lambda_{i}([S])=\frac{1}{1-(1-\lambda_{i}^{-1}(1))/[S]}.
\]

Thus, the autocorrelation decay should be slower under larger concentration.

\textup{Scenarios} 3 \textup{or} 4. The nonone eigenvalues are of order $
\tau^{-1}$, so the autocorrelation should decay extremely fast for all
concentrations.\vspace*{-2pt}
\end{theorem}

This theorem tells us that for fast fluctuation enzymes (scenarios 3 or 4),
the turnover time correlation tends to be zero. Intuitively, this is because
the fast fluctuation enzymes prefer conformation fluctuation rather than
going through the binding-association-catalytic path that leads to the
product, so in a single turnover event, the enzyme undergoes intensive
conformation changes, which effectively blurs the information on the
reaction path carried by the turnover time, resulting in zero correlation.
Under scenario 1, the autocorrelation decay pattern does not vary when the
concentration changes. This is because when the enzyme does not fluctuate,
it goes from $E_{i}$ to $\ES_{i}$ directly, and the change of concentration
consequently does not alter the distribution of the reaction path.
Thus, the
correlation between turnover times does not depend on the
concentration.\vadjust{\goodbreak}

The result from scenarios 1, 3 or 4 contradicts the experimental finding:
correlation exists between the turnover time and is stronger under higher
concentration (see Figure~\ref{fig2}). Only scenario 2 fully agrees
with the
experiments, suggesting that the enzyme--substrate complex ($\ES_{i}$) does
not fluctuate much. This is supported by recent single-molecule experimental
findings [\citet{LuXunXie98}, \citet{Yanetal03}, \citet
{Minetal05N2}] where slow
conformational fluctuation in the enzyme--substrate complexes were observed.

\subsubsection*{The effect of concentration on fluorescence intensity
autocorrelation}


We now consider the intensity autocorrelation under each of the four
scenarios. We write $\mathbf{Q}_{AA}=\mathbf{I}_{\alpha}+\mathbf
{J}_{\alpha
}$, where $\mathbf{I}_{\alpha}=\operatorname{diag}\{\alpha_{11},\ldots
,\alpha
_{nn}\}$, and $\mathbf{Q}_{BB}=\mathbf{I}_{\beta}+\mathbf{J}_{\beta}$,
where $\mathbf{I}_{\beta}=\operatorname{diag}\{\beta_{11},\ldots,\beta
_{nn}\}$%
. For scenarios 1 and 2, we assume that both the enzyme and the
enzyme--substrate complex fluctuate slowly: $\alpha_{ij}$ and $\beta
_{ij}$ $%
(i\neq j)$ are negligible, but the sums $\alpha_{ii}=-\sum_{j\neq
i}\alpha
_{ij}$ and $\beta_{ii}=-\sum_{j\neq i}\beta_{ij}$ are not.
Furthermore, we
assume that in scenario 1 the enzyme fluctuation is much slower than the
enzyme--substrate complex fluctuation (so $\mathbf{Q}_{AA}=0$ and
$\mathbf{Q}%
_{BB}=\mathbf{I}_{\beta}$ in scenario 1), and in scenario 2 the
enzyme--substrate complex fluctuation is much slower (so $\mathbf{Q}_{BB}=0$
and $\mathbf{Q}_{AA}=\mathbf{I}_{\alpha}$ scenario 2).
%
%
\begin{theorem}
\label{theo5}For enzymes with fast cycle reset, the matrix governing the
intensity autocorrelation is $\mathbf{K}$. Its eigenvalues and the
autocorrelation decay, under the four different scenarios, satisfy the
following:

\textup{Scenario} 1 ($\mathbf{Q}_{AA}=0$ and $\mathbf{Q}_{BB}=\mathbf{I}
_{\beta}$). The autocorrelation decay is slower under lower concentration,
and the dominating eigenvalues are given by%
\[
\kappa_{i}=\tfrac{1}{2}\bigl( -([S]k_{1i}-\beta_{ii} +k_{-1i}+k_{2i})+\sqrt{%
([S]k_{1i}-\beta_{ii} +k_{-1i}+k_{2i})^{2}+4\beta_{ii} [ S]k_{1i}}
\bigr).
\]

\textup{Scenario} 2 ($\mathbf{Q}_{BB}=0$ and $\mathbf{Q}_{AA}=\mathbf{I}
_{\alpha}$). The autocorrelation decay is faster under lower concentration,
and the dominating eigenvalues are given by%
\begin{eqnarray*}
\kappa_{i}&=&\tfrac{1}{2}\bigl( -([S]k_{1i}-\alpha_{ii} +k_{-1i}+k_{2i})\\
&&\hspace*{9pt}{}+\sqrt{
([S]k_{1i}-\alpha_{ii} +k_{-1i}+k_{2i})^{2}+4\alpha
_{ii}(k_{-1i}+k_{2i})}%
\bigr).
\end{eqnarray*}

\textup{Scenario} 3. The autocorrelation decay does not depend on the
concentration.

\textup{Scenario} 4. The autocorrelation decay is slower under lower
concentration.
\end{theorem}

The proof of the theorem is given in the \hyperref[app]{Appendix}. Our
results of the
dependence of turnover time autocorrelation and fluorescence intensity
autocorrelation on the substrate concentration show that in order to have
slower decay under higher substrate concentration (as seen in Figure
\ref%
{fig2}), fluctuation of both the enzyme and the enzyme--substrate complex
cannot be fast; furthermore, the fluctuation of the enzyme--substrate complex
needs to be slower than the fluctuation of the enzyme.\vadjust{\goodbreak}

In summary, each of the four scenarios yields a different autocorrelation
pattern, but only the one under scenario 2 matches the experimental finding.
Therefore, we will focus on scenario 2 from now on.

\subsection{Continuous limit}

To simplify the coefficients $\sigma_{i}$ and $C_{i}$ and to address the
number of distinct conformations $n$, we adopt the idea in the previous work
[\citet{Kouetal}, \citet{Kou08N2}] by utilizing a continuous limit.
First, we let $n\rightarrow\infty$ and in this way model the transition
rates as continuous variables with certain distributions. Consequently, we
treat the eigenvalues also as continuous variables. Second, we assume that
all the coefficients ($\sigma_{i}$~and~$C_{i}$) are proportional to the
probability weight of the conjugate eigenvalues. This assumption is partly
based on the fact that all the observed experimental correlations are
positive. With these two assumptions, the covariance can be represented
by%
\[
\operatorname{cov}(T^{1},T^{m})\propto\int\lambda^{m}f(\lambda)\,d\lambda,\qquad
\operatorname{cov}(I(0),I(t))\propto\int e^{\kappa(t-\Delta t)}g(\kappa
)\,d\kappa
,
\]
where $f$ and $g$ are the corresponding distribution functions.

$\lambda$ and $\kappa$ are functions of the transition rates. Since the
transition rates are always positive, a natural choice is to model the
transition rates as either constants or following Gamma distributions. In
the previous work [\citet{Kou08N2}] on the stochastic network
model, the
association rate $k_{1}$ and dissociation rate $k_{-1}$ are modeled as
constants while the catalytic rate $k_{2}$ follows a Gamma distribution
$%
\Gamma(a,b)$. We adopt them in our fitting.

%
%
\begin{figure}

\includegraphics{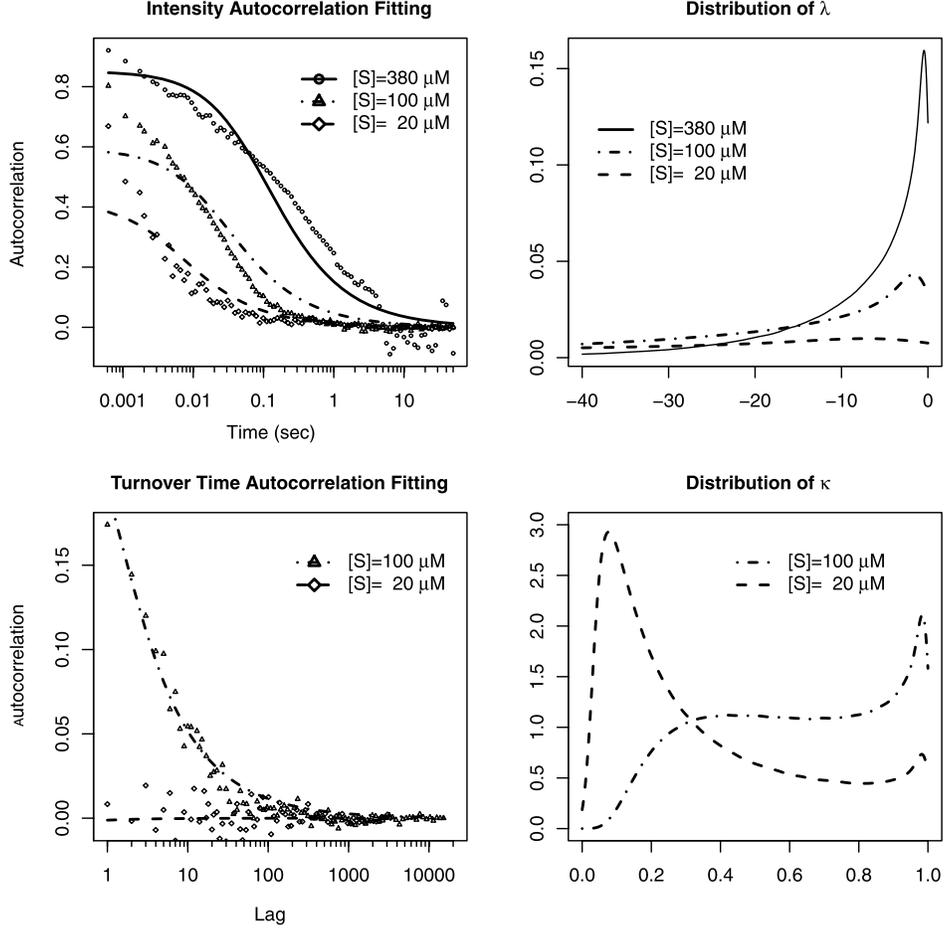}

\caption{Left: data fitting to the intensity and turnover time
autocorrelations based on~(\protect\ref{lam}) and (\protect
\ref{kappa}). Right: the corresponding distributions of the eigenvalues
$\protect\lambda$ and $\protect\kappa$.}
\label{fig3}\vspace*{-2pt}
\end{figure}

We know from Section~\ref{sec41} that scenario 2 matches the experimental
finding, so we take $\mathbf{Q}_{BB}=0$ and $\mathbf{Q}_{AA}=\mathbf{I}%
_{\alpha}$. Then the eigenvalue $\lambda$ (based on Theorem~\ref{theo4}
and its proof in the \hyperref[app]{Appendix}) is given by%
%
%
\begin{equation}\label{lam}
\lambda=\frac{1}{1+\alpha^{\ast}(k_{-1}+k_{2})/([S]k_{1}k_{2})},
\end{equation}
and the eigenvalue $\kappa$ (based on Theorem~\ref{theo5}) is%
%
%
\begin{eqnarray} \label{kappa}
\kappa&=&\tfrac{1}{2}\bigl(
-([S]k_{1}+\alpha^{\ast}+k_{-1}+k_{2})\nonumber\\[-8pt]\\[-8pt]
&&\hspace*{9pt}{}+\sqrt{%
([S]k_{1}+\alpha^{\ast}+k_{-1}+k_{2})^{2}-4\alpha^{\ast
}(k_{-1}+k_{2})}%
\bigr),\nonumber
\end{eqnarray}
where $\alpha^{\ast}$ stands for a generic $-\alpha_{ii}$ (since we are
taking the continuous version). For the distribution of $\alpha^{\ast}$
(i.e., the distribution of $-\alpha_{ii}$), we note that, first, its support
should be the positive real line, and, second, $-\alpha_{ii}=\sum
_{j\neq
i}\alpha_{ij}$ is a sum of many random variables $\alpha_{ij}$ from a
common distribution, so we expect that the distribution of $\alpha
^{\ast}$
should be infinitely divisible. These two considerations lead us to
assume a
Gamma distribution $\Gamma(a_{\alpha},b_{\alpha})$ for~$\alpha
^{\ast}$.

\subsection{Data fitting}

The data available to us include the intensity correlation under three
concentrations: $[S]=$ 380, 100 and 20 $\mu M$ (micro molar), and turnover
time autocorrelation under two concentrations\vadjust{\goodbreak} $[S]=$ 100 and 20 $\mu
M$. We
calculated eigenvalues based on~(\ref{lam}) and~(\ref{kappa}),
where $k_{1}$ and $k_{-1}$ are constants, $\alpha^{\ast}$~and $k_{2}$
follow distributions $\Gamma(a_{\alpha},b_{\alpha})$ and $\Gamma(a,b)$,
respectively. The parameters of interests are
$k_{1},k_{-1},a,b,a_{\alpha}$
and $b_{\alpha}$. The best fits are found through minimizing the square
distance between the theoretical and observed values. The parameters are
estimated as follows: $k_{1}=1.785\times10^{3}$ $(\mu M)^{-1}s^{-1}$, $%
k_{-1}=6.170\times10^{3}$ $s^{-1}$, $a=13.49$, $b=2.279$ $s^{-1}$, $a_{\alpha
}=0.6489$, and $b_{\alpha}=1.461\times10^{3}$ $s^{-1}$ ($s$ stands for
second). Figure~\ref{fig3} shows the fitting of the autocorrelation
functions and the distributions of the eigenvalues.

Figure~\ref{fig3} shows that our model gives a good fit to the turnover time
autocorrelation and an adequate fit to the fluorescence intensity
autocorrelation, capturing\vadjust{\goodbreak} the main trend in the intensity autocorrelation.
The distributions of the eigenvalues in the right panels clearly indicate
that higher substrate concentration corresponds to larger eigenvalues, which
are then responsible for the slower decay of the autocorrelations.

Our model thus offers an adequate explanation of the observed decay patterns
of the autocorrelation functions. The stochastic network model tells us why
the decay must be multi-exponential. It further explains why the decay is
slower under higher substrate concentration. Our consideration of the
different scenarios also provides insight on the enzyme's conformational
fluctuation: slow fluctuation, particularly of the enzyme--substrate complex,
gives rise to the experimentally observed autocorrelation decay pattern.

\section{Discussion}
\label{sec5}

In this article we explored the stochastic network model previously
developed to account for the empirical puzzles arising from recent
single-molecule enzyme experiments. We conducted a detailed study of the
autocorrelation function of the turnover time and of the fluorescence
intensity and investigate the effect of substrate concentration on the
correlations.

Our analytical results show that (a) the stochastic network model gives
multi-exponential autocorrelation decay of both the turnover times and
the fluorescence intensity, agreeing with the experimental observation;
(b) under suitable conditions, the autocorrelation decays more slowly
with higher concentration, also agreeing with the experimental result;
(c) the slower autocorrelation decay under higher concentration implies
that the fluctuation of the enzyme--substrate complex should be slow,
corroborating the conclusion from other single-molecule experiments
[\citet{LuXunXie98}, \citet{Yanetal03},
\citet{Minetal05N2}]. In addition to providing a theoretical
underpinning of the experimental observations, the numerical result
from the
model fits well with the experimental autocorrelation as seen in
Section \ref
{sec4}.

Some problems remain open for future investigation:\vspace*{8pt}

(1) When we discussed the dependence of intensity autocorrelation on
substrate concentration in Section~\ref{sec41}, we approximated the
fluctuation transition matrix with its diagonal entries. This simple
approximation provides useful insight into the decay pattern under different
concentration. A better approximation that goes beyond the diagonal entries
is desirable. It might lead to a better fitting to the experimental data.

(2) We used Gamma distribution to model the transition rates. This is purely
statistical. Can it be derived from a physical angle? If so, the connection
not only will lead to better estimation, but also provides new insight into
the underlying mechanism of the enzyme's conformation fluctuation.

(3) We used the continuous limit $n\rightarrow\infty$ to do the data
fitting so that the number of parameters reduces from more than $3n$ to a
manageable six. Obtaining the standard error for the estimates is open for
future investigation. The main difficulties are the lack of tractable tools
to approximate the standard\vadjust{\goodbreak} error of the autocorrelation estimates and the
challenge to carry out a Monte Carlo estimate ($n$ needs to be quite large
for an ad hoc Monte Carlo simulation, but such an $n$ will bring back a
large number of unspecified parameters).

Single-molecule biophysics, like many newly emerging fields, is
interdisciplinary. It lies at the intersection of biology, chemistry and
physics. Owing to the stochastic nature of the nano world, single-molecule
biophysics also presents statisticians with new problems and new challenges.
The stochastic model for single-enzyme reaction represents only one such
case among many interesting opportunities. We hope this article will
generate further interest in solving biophysical problems with modern
statistical methods; and we believe that the knowledge and tools gained in
this process will in turn advance the development of statistics and
probability.

\begin{appendix}\label{app}
\section*{Appendix: Proofs}

\begin{pf*}{Proof of Lemma~\ref{lem4}} Using the first-step analysis, we
have%
%
\[
\mathbf{G}%
\pmatrix{
\mathbf{P}_{AC} \cr
\mathbf{P}_{BC}%
}
=-%
\pmatrix{
\mathbf{0} \cr
\mathbf{Q}_{BC}%
},
\]
where%
\[
\mathbf{G}=\pmatrix{
\mathbf{Q}_{AA}-\mathbf{Q}_{AB} & \mathbf{Q}_{AB} \cr
\mathbf{Q}_{BA} & \mathbf{Q}_{BB}-(\mathbf{Q}_{BA}+\mathbf{Q}_{BC})}
=
\pmatrix{
\mathbf{L} & \mathbf{M} \cr
\mathbf{N} & \mathbf{R}}^{-1}.
\]
Only the diagonal elements of $\mathbf{G}$ are negative, and its row
sums are either 0 or negative. Thus, $-\mathbf{G}$ is a stable matrix
[\citet{HorJoh85}], which always has an inverse. Thus, we have
%
\[
\pmatrix{
\mathbf{P}_{AC} \cr
\mathbf{P}_{BC}}
=-\mathbf{G}^{-1}%
\pmatrix{
\mathbf{0} \cr
\mathbf{Q}_{BC}}
=%
\pmatrix{
-\mathbf{MQ}_{BC} \cr
-\mathbf{RQ}_{BC}}.
\]
For $P_{E_{i}^{0}E_{j}}$, similarly, we have%
%
\[
\mathbf{P}_{CA}=-(\mathbf{Q}_{CC}-\mathbf{Q}_{CA})^{-1}\mathbf{Q}_{CA}=(
\mathbf{I}-\mathbf{Q}_{CA}^{-1}\mathbf{Q}_{CC})^{-1}.
\]
\upqed
\end{pf*}
\begin{pf*}{Proof of Lemma~\ref{lem5}} For $E(T_{E_{i}E_{j}^{0}})$,
when the
first-step analysis is applied, the first-step probability should be
conditioned on the exit state~$E_{j}^{0}$, that is, $P(E_{i}$ returns
to $E_{k}$
first $|$ exit at $E_{j}^{0})=P(E_{i}$ returns to $E_{k}$ first$)P_{E_{k}E_{j}^{0}}/\break P_{E_{i}E_{j}^{0}}$. Thus, we have the following
equation:
\begin{eqnarray*}
E(T_{E_{i}E_{j}^{0}})&=&\biggl( 1+k_{1i}[S]\frac{P_{\ES_{i}E_{j}^{0}}}{%
P_{E_{i}E_{j}^{0}}}E(T_{\ES_{i}E_{j}^{0}})+\sum_{k\neq i}\alpha
_{ik}\frac{%
P_{E_{k}E_{j}^{0}}}{P_{E_{i}E_{j}^{0}}}E(T_{E_{k}E_{j}^{0}})\biggr)\\
&&{}\bigg /\biggl(
k_{1i}[S]+\sum_{k\neq i}\alpha_{ik}\biggr) .
\end{eqnarray*}
Similar expression can be derived for $E(T_{\ES_{i}E_{j}^{0}})$.
Together we
have%
\[
\pmatrix{
\mathbf{E}_{AC} \cr
\mathbf{E}_{BC}}
=-\mathbf{G}^{-1}%
\pmatrix{
\mathbf{P}_{AC} \cr
\mathbf{P}_{BC}}
=\pmatrix{
\mathbf{LMQ}_{BC}+\mathbf{MRQ}_{BC} \cr
\mathbf{NMQ}_{BC}+\mathbf{RRQ}_{BC}}
.
\]
\upqed
\end{pf*}
\begin{pf*}{Proof of Theorem~\ref{theo1}} $\operatorname{cov}%
(T^{1},T^{m})=E(T^{1}T^{m})-E(T^{1})E(T^{m})$. The first term $E(T^{1}T^{m})$
can be expressed as%
\[
E(T^{1}T^{m})=%
\sum
_{i,j,k,l}w(E_{i})P_{E_{i}E_{j}^{0}}E(T_{E_{i}E_{j}^{0}})P_{E_{j}^{0}E_{k}}P_{E_{k}E_{l}}^{(m-2)}E(T_{E_{l}}),
\]
that is, the system starts the first turnover event from $E_{i}$, ends
it in
$E_{j}^{0}$, then starts the second from $E_{k}$, repeats this
procedure for
$m-2$ times, and finally starts the last turnover from $E_{l}$. Note
that $%
[P_{E_{k}E_{l}}^{(m-2)}]_{n\times n}=(\mathbf{P}_{AC}\mathbf
{P}_{CA})^{m-2}$%
. Thus,\vspace*{1pt} using the matrices defined in Lemmas~\ref{lem1} to~\ref{lem5}, we
have%
\[
\operatorname{cov}(T^{1},T^{m})=\mathbf{wE}_{AC}\mathbf{P}_{CA}(\mathbf
{P}_{AC}%
\mathbf{P}_{CA})^{m-2}\bolds{\mu}_{A}-(\mathbf{w}\bolds{\mu}_{A})^{2}.
\]
%
Applying\vspace*{1pt} the results of Lemmas~\ref{lem4} and~\ref{lem5} and the facts
that $%
\mathbf{R}=\break(\mathbf{I}-\mathbf{Q}_{AB}^{-1}\mathbf{Q}_{AA})\mathbf{M}$
and $(%
\mathbf{w}\bolds{\mu}_{A})^{2}=-\mathbf{w(\mathbf{L}+\mathbf{M})%
\mathbf{1}w}\bolds{\mu}_{A}=-\mathbf{w}(
\mathbf{L}+\mathbf{M%
}(\mathbf{I}-\break\mathbf{Q}_{AB}^{-1}\mathbf{Q}_{AA}))
\times\mathbf{1w}\bolds{\mu}_{A}$, we can finally arrange the covariance
as%
%
\[
\operatorname{cov}(T^{1},T^{m})=-\mathbf{w}\bigl(\mathbf{L}+\mathbf
{M}(\mathbf{I}-%
\mathbf{Q}_{AB}^{-1}\mathbf{Q}_{AA})\bigr)[(\mathbf{P}_{AC}\mathbf
{P}_{CA})^{m-1}-%
\mathbf{1w}]\bolds{\mu}_{A}.
\]
\upqed
\end{pf*}
\begin{pf*}{Proof of Corollary~\ref{cor1}} We only need to prove that $%
\mathbf{1}$ and $\mathbf{w}$ are, respectively, the right and left
eigenvectors of $\mathbf{P}_{AC}\mathbf{P}_{CA}$ associated with
the~eigenvalue 1. The first is a direct consequence of the fact that
$\mathbf{P}%
_{AC}\mathbf{P}_{CA}$ is~a~stochastic matrix. The second can be
verified by
observing that\break $-\mathbf{wMQ}_{BC}(\mathbf{I}-\mathbf
{Q}_{CA}^{-1}\mathbf{Q}%
_{CC})^{-1}=\mathbf{w}$ through~(\ref{eqV}).
\end{pf*}
\begin{pf*}{Proof of Theorem~\ref{theo2}} In~(\ref{I}), the second
term $N_{t}^{0}(\Delta t)$ represents the independent background noise during
period $(t-\Delta t,t) $. Thus,%
\begin{eqnarray*}
\operatorname{cov}(I(0),I(t))
&=&E[N_{t}(T_{\mathrm{on}}(t))N_{0}(T_{\mathrm{on}}(0))]-E[N_{t}(T_{\mathrm{on}}(t))]E[N_{0}(T_{\mathrm{on}}(0))]
\\
&=&\nu^{2}[E(T_{\mathrm{on}}(t)T_{\mathrm{on}}(0))-E(T_{\mathrm{on}}(t))E(T_{\mathrm{on}}(0))].
\end{eqnarray*}
Let $\mathcal{S}=\{E_{1},\ldots,E_{n},\ES_{1},\ldots
,\ES_{n},E_{1}^{0},\ldots,E_{n}^{0}\}$ be the set of all possible states.
Let $X_{t}$ be the process evolving according to~(\ref{2by2}). Let $\pi_{i}$
be the equilibrium probability of state $i$ and $P_{ij}(s)$ be the
transition probability from state $i$ to state $j$ after time $s$. We
have%
\begin{eqnarray*}
&&
E(T_{\mathrm{on}}(t)T_{\mathrm{on}}(0))-E(T_{\mathrm{on}}(t))E(T_{\mathrm{on}}(0)) \\
&&\qquad=\sum_{i,j,k,l\in\mathcal{S}}\pi_{i}P_{ij}(\Delta t)P_{jk}(t-\Delta
t)P_{kl}(\Delta t)E\bigl(T_{\mathrm{on}}(0)|X_{-\Delta
t}=i,X_{0}=j\bigr)\\
&&\hspace*{64pt}{}\times E\bigl(T_{\mathrm{on}}(t)|X_{t-\Delta t}=k,X_{t}=l\bigr) \\
&&\qquad\quad{}-\sum_{i,j,k,l\in\mathcal{S}}\pi_{i}P_{ij}(\Delta t)\pi
_{k}P_{kl}(\Delta t)E\bigl(T_{\mathrm{on}}(0)|X_{-\Delta
t}=i,X_{0}=j\bigr)\\[-2pt]
&&\hspace*{78pt}{}\times
E\bigl(T_{\mathrm{on}}(t)|X_{t-\Delta t}=k,X_{t}=l\bigr) \\[-2pt]
&&\qquad=\sum_{i,j,k,l\in\mathcal{S}}\pi_{i}P_{ij}(\Delta t)P_{kl}(\Delta
t)E\bigl(T_{\mathrm{on}}(0)|X_{-\Delta t}=i,X_{0}=j\bigr) \\[-2pt]
&&\hspace*{32pt}\qquad\quad{}\times E\bigl(T_{\mathrm{on}}(t)|X_{t-\Delta t}=k,X_{t}=l\bigr)\{ P_{jk}(t-\Delta t)-\pi
_{k}\}.
\end{eqnarray*}

The probability transition matrix $[P_{ij}(t)]_{3n\times3n}$ is the matrix
exponential of the generating matrix~(\ref{2by2}):
$[P_{ij}(t)]_{3n\times
3n}=\exp(\mathbf{Q}t)$. Zero is an eigenvalue of $\mathbf{Q}$ with right
eigenvector $\mathbf{1}$ and left eigenvector $\bolds{\pi}$, the stationary
distribution. Assume $\mathbf{Q}$ is diagonalizable. Let $\mu_{i}$, $%
i=2,3,\ldots,3n$, denote the other eigenvalues, and $\bolds{\xi}_{i}$
and $%
\bolds{\eta}_{i}^{T}$ be the corresponding right and left eigenvectors.
We have
\[
\exp(\mathbf{Q}t)=\mathbf{1}\bolds{\pi}+\sum_{i=2}^{3n}e^{\mu
_{i}t}\bolds
{\xi}%
_{i}\bolds{\eta}_{i}^{T}.
\]
Therefore, we can rewrite%
\[
\operatorname{cov}(I(0),I(t))\propto\sum_{i=2}^{3n}C_{i}e^{\mu
_{i}(t-\Delta
t)}.\vspace*{-2pt}
\]
\upqed
\end{pf*}

To prove Theorem~\ref{theo3}, we need the following two useful lemmas
on the
eigenvalues of a matrix.\vspace*{-2pt}
%
%
\begin{lemma}[{[Theorems 6.1.1 and 6.4.1 of \citet{HorJoh85}]}]
\label{lemA} Let $%
A=[a_{ij}]\in M_{n}$, where $M_{n}$ is the set of all complex matrices.
Let $%
\alpha\in[0,1]$ be given and define $R_{i}^{\prime}$ and $%
C_{i}^{\prime}$ as the deleted row and column sums of $A$,
respectively,%
\[
R_{i}^{\prime}=\sum_{j\neq i}|a_{ij}|,\qquad C_{i}^{\prime
}=\sum_{j\neq i}|a_{ji}|.
\]
Then, (1) all the eigenvalues of $A$ are located in the union of $n$
discs%
%
%
\begin{equation} \label{radius}
\bigcup_{i=1}^{n}\{z\in\mathbf{C}\dvtx|z-a_{ii}|\leq R_{i}^{\prime\alpha
}C_{i}^{\prime1-\alpha}\}.
\end{equation}
(2) Furthermore, if a union of $k$ of these $n$ discs forms a connected
region that is disjoint from all the remaining $n-k$ discs, then there are
precisely~$k$ eigenvalues of $A$ in this region.\vspace*{-2pt}
\end{lemma}
%
%
\begin{lemma}[{[pages 63--67 of \citet{Wil88}]}]
\label{lemB} Let $A$ and $B$ be matrices with
elements satisfying $|a_{ij}|<1,|b_{ij}|<1$. If $\lambda_{1}$ is a simple
eigenvalue (i.e., an eigenvalue with multiplicity 1) of $A$, then for matrix
$A+\varepsilon B$, where $\varepsilon$ is sufficiently small, there
will be a
eigenvalue $\lambda_{1}(\varepsilon)$ of $A+\varepsilon B$ such that%
\[
|\lambda_{1}(\varepsilon)-\lambda_{1}|=O(\varepsilon).\vadjust{\goodbreak}
\]
Furthermore, if we know that one eigenvector of $A$ associated with
$\lambda
_{1}$ is $\mathbf{x}_{1}$, then there is an eigenvector $\mathbf{x}%
_{1}(\varepsilon)$ of $A+\varepsilon B$ associated with $\lambda
_{1}(\varepsilon)$
such that%
\[
|\mathbf{x}_{1}(\bolds\varepsilon)-\mathbf{x}_{1}|=O(\varepsilon).
\]
\end{lemma}

Note that since dividing a matrix by a constant only changes the eigenvalues
with the same proportion, the condition that the entries of $A$ and $B$ are
bounded by 1 can be relaxed to that the entries of $A$ and $B$ are bounded
by a finite positive number.
\begin{pf*}{Proof of Theorem~\ref{theo3}} According to Lemma \ref
{lemA}, all
the eigenvalues of $\mathbf{Q}$ must lie in the union of discs centered
at $%
Q_{ii}$ with radii defined by~(\ref{radius}). If we take $\alpha=1/2$
in (%
\ref{radius}), then the first $n$ discs corresponding to
the diagonal entries of $\mathbf{Q}_{AA}-\mathbf{Q}_{AB}$ have centers
$O(1)$ and
radii $%
O(\delta^{1/2})$; the second $n$ discs corresponding\vspace*{1pt} to the diagonal
entries of $\mathbf{Q}_{BB}-\mathbf{Q}_{BC}-\mathbf{Q}_{BA}$ have
centers $%
O(1)$ and radii $O(1)$; the third $n$ discs corresponding to the diagonal
entries of $\mathbf{Q}_{CC}-\mathbf{Q}_{CA}$ have centers $O(\delta)$ and
radii $O(\delta^{1/2})$. Thus, for $\delta$ large enough, the union
of the
first $2n$ discs does not overlap with the union of the last $n$ discs, so
we know from Lemma~\ref{lemA} that $\mathbf{Q}$ has $2n$ eigenvalues with
order $O(\delta^{1/2})$ in the union of the first $2n$ discs and $n$ other
eigenvalues with order $O(\delta)$ in the union of the last $n$ discs.

For the $n$ eigenvalues with order $O(\delta)$, consider the following two
matrices:%
\begin{eqnarray*}
\mathbf{Y}&=&\left[
\matrix{
\mathbf{0} & \mathbf{0} & \mathbf{0} \cr
\mathbf{0} & \mathbf{0} & \mathbf{0} \vspace*{2pt}\cr
\dfrac{1}{\delta}\mathbf{Q}_{CA} & \mathbf{0} & -\dfrac{1}{\delta
}\mathbf{Q}
_{CA}}
\right], \\
\mathbf{Z}&=&\left[
\matrix{
\mathbf{Q}_{AA}-\mathbf{Q}_{AB} & \mathbf{Q}_{AB} & \mathbf{0} \cr
\mathbf{Q}_{BA} & \mathbf{Q}_{BB}-\mathbf{Q}_{BA}-\mathbf{Q}_{BC} &
\mathbf{Q%
}_{BC} \cr
\mathbf{0} & \mathbf{0} & \mathbf{Q}_{CC}}
\right] .
\end{eqnarray*}
We have $\frac{1}{\delta}\mathbf{Q}=\mathbf{Y}+\frac{1}{\delta}\mathbf
{Z}$%
. Zero is an eigenvalue of $\mathbf{Y}$ with multiplicity $2n$, and the
other $n$ eigenvalues of $\mathbf{Y}$ are $-q_{1},-q_{2},\ldots,-q_{n}$.
For large\vspace*{1pt} $\delta$, according to Lemma~\ref{lemB}, there exists $n$
eigenvalues of $\frac{1}{\delta}\mathbf{Q}$ that satisfy%
\[
\mu_{2n+i}/\delta=-q_{i}+O(\delta^{-1}),\qquad i=1,2,\ldots,n,
\]
that is,%
\[
|\mu_{i}+\delta q_{i-2n}|=\delta O(\delta^{-1})=O(1),\qquad
i=2n+1,2n+2,\ldots,3n.
\]
Now for the $2n-1$ nonzero eigenvalues of $\mathbf{Q}$ with order
$O(\delta
^{1/2})$, they are the solutions of%
\[
\vert\mathbf{Q}-\mu_{i}\mathbf{I}_{3n}\vert=0,\qquad i=2,\ldots
,2n.
\]
For large $\delta$, the matrix $\mathbf{Q}_{CC}-\mathbf{Q}_{CA}-\mu
_{i}%
\mathbf{I}_{n}$ is invertible, since it is strictly diagonal dominated. We
can decompose the determinant as%
\[
\vert\mathbf{Q}-\mu_{i}\mathbf{I}_{3n}\vert=\vert\mathbf{%
U}(\mu_{i})\vert\vert\mathbf{Q}_{CC}-\mathbf{Q}_{CA}-\mu_{i}%
\mathbf{I}_{n}\vert=0,\vadjust{\goodbreak}
\]
where%
\[
\mathbf{U}(\mu_{i})=\left[
\matrix{
\mathbf{Q}_{AA}-\mathbf{Q}_{AB}-\mu_{i}\mathbf{I}_{n} & \mathbf{Q}_{AB}
\cr
\mathbf{Q}_{BA}-\mathbf{Q}_{BC}(\mathbf{Q}_{CC}-\mathbf{Q}_{CA}-\mu_{i}
\mathbf{I}_{n})^{-1}\mathbf{Q}_{CA} & \mathbf{Q}_{BB}-\mathbf{Q}_{BA}\cr
& \hphantom{0}\qquad{}-\mathbf{Q}_{BC}-\mu_{i}\mathbf{I}_{n}}
\right] .
\]
Therefore, $\mu_{i}$, $i=2,\ldots,2n$, is also the eigenvalue of the
matrix%
\[
\left[
\matrix{
\mathbf{Q}_{AA}-\mathbf{Q}_{AB} & \mathbf{Q}_{AB} \cr
\mathbf{Q}_{BA}-\mathbf{Q}_{BC}(\mathbf{Q}_{CC}-\mathbf{Q}_{CA}-\mu_{i}
\mathbf{I}_{n})^{-1}\mathbf{Q}_{CA} & \mathbf{Q}_{BB}-\mathbf{Q}_{BA}-%
\mathbf{Q}_{BC}}
\right] =\mathbf{K}+\mathbf{S}
\]
with%
\[
\mathbf{S}=\left[
\matrix{
\mathbf{0} & \mathbf{0} \cr
-\mathbf{Q}_{BC}-\mathbf{Q}_{BC}(\mathbf{Q}_{CC}-\mathbf{Q}_{CA}-\mu
_{i}%
\mathbf{I}_{n})^{-1}\mathbf{Q}_{CA} & \mathbf{0}}
\right] .
\]
We note that $\mathbf{I}_{n}+\mathbf{Q}_{BC}(\mathbf{Q}_{CC}-\mathbf{Q}%
_{CA}-\mu_{i}\mathbf{I}_{n})^{-1}\mathbf{Q}_{CA}=(\mathbf{W}-\mathbf{I}
_{n})^{-1}\mathbf{W}$, where $\mathbf{W}=\mathbf{Q}_{CA}^{-1}(\mathbf{Q}
_{CC}-\mu_{i}\mathbf{I}_{n})$. Since\vspace*{1pt} $\mathbf{Q}_{CA}$ is of the order
$%
O(\delta)$ and $\mu_{i}$ is of the order $O(\delta^{1/2})$, the entries
of $\mathbf{W}$ are of the order $O(\delta^{-1/2})$, so are the
entries of $%
\mathbf{S}$. Applying Lemma~\ref{lemB} to $\mathbf{K}+\mathbf{S}$ tells us
that for each $\mu_{i}$ there must be an eigenvalue $\kappa_{i}$ of $%
\mathbf{K}$, which has the property that%
\[
\mu_{i}=\kappa_{i}+O(\delta^{-1/2}),\qquad i=2,\ldots,2n.
\]
\upqed\end{pf*}
\begin{pf*}{Proof of Theorem~\ref{theo4}} We know from Lemma~\ref{lem1} that
\[
\mathbf{M}=[ \mathbf{Q}_{BB}-\mathbf{Q}_{BC}-(\mathbf{Q}_{BB}-\mathbf{Q}
_{BA}-\mathbf{Q}_{BC})\mathbf{Q}_{AB}^{-1}\mathbf{Q}_{AA}] ^{-1}.
\]

\textit{Scenario} 1. When $\mathbf{Q}_{AA}=0$, $\mathbf{M}=(\mathbf
{Q}_{BB}-%
\mathbf{Q}_{BC})^{-1}$, so the eigenvalues and eigenvectors of $-\mathbf
{MQ}%
_{BC}$ have nothing to do with $[S]$.

\textit{Scenario} 2. When\vspace*{1pt} $\mathbf{Q}_{BB}=0$, $(-\mathbf
{MQ}_{BC})^{-1}=%
\mathbf{I}_{n}-\frac{1}{[S]}\mathbf{Q}_{BC}^{-1}(\mathbf{Q}_{BA}+\break\mathbf
{Q}%
_{BC})\mathbf{\tilde{Q}}_{AB}^{-1}\times\mathbf{Q}_{AA}$. Thus, if $-\mathbf
{MQ}%
_{BC}$ has eigenvalue $\lambda_{i}(1)$ when $[S]=1$, then for general
$[S]$%
, $-\mathbf{MQ}_{BC}$ has eigenvalue
\[
\lambda_{i}([S])=\frac{1}{1-(1-\lambda_{i}^{-1}(1))/[S]}.
\]

\textit{Scenario} 3. We write $\mathbf{Q}_{AA}=\tau\mathbf{\tilde
{Q}}_{AA}$%
, where $\tau$ is large. Then $-\mathbf{MQ}_{BC}$ is%
\begin{eqnarray*}
(-\mathbf{MQ}_{BC})^{-1}& = &\mathbf{I}-\mathbf{Q}_{BC}^{-1}\mathbf
{Q}_{BB}+%
\mathbf{Q}_{BC}^{-1}(\mathbf{Q}_{BB}-\mathbf{Q}_{BA}-\mathbf
{Q}_{BC})\mathbf{%
Q}_{AB}^{-1}\mathbf{\tilde{Q}}_{AA}\cdot\tau\\
& = &\tau\biggl[\frac{1}{\tau}(\mathbf{I}-\mathbf{Q}_{BC}^{-1}\mathbf
{Q}%
_{BB})+\mathbf{Q}_{BC}^{-1}(\mathbf{Q}_{BB}-\mathbf{Q}_{BA}-\mathbf
{Q}_{BC})%
\mathbf{Q}_{AB}^{-1}\mathbf{\tilde{Q}}_{AA}\biggr].
\end{eqnarray*}
Suppose the eigenvalues of $\mathbf{Q}_{BC}^{-1}(\mathbf{Q}_{BB}-\mathbf
{Q}%
_{BA}-\mathbf{Q}_{BC})\mathbf{Q}_{AB}^{-1}\mathbf{\tilde{Q}}_{AA}$ are $
0,\lambda_{2}^{\ast},\ldots,\allowbreak\lambda_{n}^{\ast}$. Then according to
Lemma~\ref{lemB}, the eigenvalues of $\frac{1}{\tau}(\mathbf
{I}_{n}-\mathbf{%
Q}_{BC}^{-1}\mathbf{Q}_{BB})+(\mathbf{Q}_{BB}-\mathbf{Q}_{BA}-\mathbf{Q}
_{BC})\mathbf{Q}_{AB}^{-1}\mathbf{\tilde{Q}}_{AA}$ are%
\[
O(\tau^{-1}),\lambda_{2}^{\ast}+O(\tau^{-1}),\ldots,\lambda
_{n}^{\ast
}+O(\tau^{-1}).
\]
Thus, the eigenvalues of $-\mathbf{MQ}_{BC}$ are%
\[
1,\frac{1}{\tau\lambda_{2}^{\ast}+O(1)},\ldots,\frac{1}{\tau\lambda
_{n}^{\ast}+O(1)},
\]
namely, all the nonone eigenvalues of $-\mathbf{MQ}_{BC}$ are of order $
\tau^{-1}$.

\textit{Scenario} 4. Using an identical method as in scenario 3, we can show
that all the nonone eigenvalues of $-\mathbf{MQ}_{BC}$ are of order
$\tau
^{-1}$.
\end{pf*}
\begin{pf*}{Proof of Theorem~\ref{theo5}} The matrix $\mathbf{K}$ can be
written as%
\begin{eqnarray*}
\mathbf{K} &=&\pmatrix{
\mathbf{I}_{\alpha}-\mathbf{Q}_{AB} & \mathbf{Q}_{AB} \cr
\mathbf{Q}_{BA}+\mathbf{Q}_{BC} & \mathbf{I}_{\beta}-\mathbf{Q}_{BA}-%
\mathbf{Q}_{BC}}
+\pmatrix{
\mathbf{J}_{\alpha} & \mathbf{0} \cr
\mathbf{0} & \mathbf{J}_{\beta}}
\\
&=&\mathbf{T}+\pmatrix{
\mathbf{J}_{\alpha} & \mathbf{0} \cr
\mathbf{0} & \mathbf{J}_{\beta}}.
\end{eqnarray*}
We thus know from Lemma~\ref{lemB} that the eigenvalues of $\mathbf{K}$ can
be approximated by the eigenvalues of $\mathbf{T}$. If $\vert\mathbf{I}
_{\alpha}-\mathbf{Q}_{AB}-\kappa\mathbf{I}_{n}\vert$ is invertible,
then%
\begin{eqnarray*}
\vert\mathbf{T}-\kappa\mathbf{I}_{n}\vert&=&\vert\mathbf{%
I}_{\alpha}-\mathbf{Q}_{AB}-\kappa\mathbf{I}_{n}\vert\\
&&{}\times\vert\mathbf{I}_{\beta}-\mathbf{Q}_{BA}-\mathbf{Q}_{BC}\\
&&\hspace*{14pt}{}+(%
\mathbf{Q}_{BA}+\mathbf{Q}_{BC})(-\mathbf{I}_{\alpha}+\mathbf{Q}%
_{AB}+\kappa\mathbf{I}_{n})^{-1}\mathbf{Q}_{AB}-\kappa\mathbf{I}%
_{n}\vert;
\end{eqnarray*}
we know that any eigenvalue $\kappa$ of $\mathbf{T}$ must make the second
determinant on the right-hand side zero. This determinant only involves
diagonal matrices, so we have%
%
%
\begin{equation} \label{root}
\kappa-\beta_{ii}+k_{2i}+k_{-1i}=\frac
{[S]k_{1i}(k_{2i}+k_{-1i})}{\kappa
-\alpha_{ii}+[S]k_{1}}.
\end{equation}
If $\vert\mathbf{I}_{\alpha}-\mathbf{Q}_{AB}-\kappa\mathbf{I}%
_{n}\vert$ is not invertible, then there is at least one $j$ so that $%
\kappa=\alpha_{jj}-k_{1j}[S]$. But it can be verified that in order to
make $\kappa$ an eigenvalue of~$\mathbf{T}$, there must exit another
$i\neq
j$ such that~(\ref{root}) holds for this $\kappa$. Therefore, any
eigenvalue must be a root of~(\ref{root}).

Equation~(\ref{root}) has two negative roots for each $i$, but we only need to
consider the root closer to 0, since it dominates the decay.

\textit{Scenario} 1. $\alpha_{ii} =0$. The root is%
\[
\kappa_{i}=\tfrac{1}{2}\bigl( -([S]k_{1i}-\beta_{ii} +k_{-1i}+k_{2i})+\sqrt{%
([S]k_{1}-\beta_{ii} +k_{-1i}+k_{2i})^{2}+4\beta_{ii} [ S]k_{1}}%
\bigr),
\]
which is monotone decreasing in $[S]$.

\textit{Scenario} 2. $\beta_{ii}=0$. The root is%
\begin{eqnarray*}
\kappa_{i}&=&\tfrac{1}{2}\bigl( -([S]k_{1i}-\alpha_{ii}+k_{-1i}+k_{2i})\\
&&\hspace*{9pt}{}+\sqrt{
([S]k_{1}-\alpha_{ii}+k_{-1i}+k_{2i})^{2}+4\alpha
_{ii}(k_{-1i}+k_{2i})}%
\bigr),
\end{eqnarray*}
which is monotone increasing in $[S]$.

\textit{Scenario} 3. $\mathbf{Q}_{AA}=\tau\mathbf{\tilde{Q}}_{AA}$,
where $%
\tau$ is large. Following the same method as we used in the proof of
Theorem~\ref{theo3}, we can show that $n$ eigenvalues of $\mathbf{K}$
are of
the order $O(\tau)$ and they will not contribute much to the correlation.
The other eigenvalues governing the decay pattern can be approximated
by the
eigenvalues of $\mathbf{Q}_{BB}-\mathbf{Q}_{BA}-\mathbf{Q}_{BC}$, which do
not depend on the concentration $[S]$.

\textit{Scenario} 4. $\mathbf{Q}_{BB}=\tau\mathbf{\tilde{Q}}_{BB}$,
where $\tau$ is large. Using the same method as in the proof of Theorem \ref{theo3},
we can show that the dominating eigenvalues of~$\mathbf{K}$ can be
approximately by the eigenvalues of $\mathbf{Q}_{AA}-\mathbf{Q}_{AB}$. Since
we know that $\mathbf{Q}_{AA}\approx\mathbf{I}_{\alpha}$, the eigenvalues
of $\mathbf{Q}_{AA}-\mathbf{Q}_{AB}$ is approximately $\alpha
_{ii}-[S]k_{1i} $, which is monotone decreasing in $[S]$.
\end{pf*}
\end{appendix}

\section*{Acknowledgments}

The authors thank the Xie group at the Department of Chemistry and
Chemical Biology of Harvard University for sharing the experimental
data.


%

\printaddresses

\end{document}